\documentstyle[11pt,paspconf,epsf]{article}

\begin {document}

\title{Searches for Quasars at $z > 5$}

\author{Patrick S. Osmer}

\affil{Department of Astronomy, The Ohio State University, 
140. W. 18th Ave., Columbus, OH 43210 USA, 
posmer@astronomy.ohio-state.edu}

\begin{abstract}

Quasars continue to be the most luminous objects known in the
universe but no longer have the largest redshifts\footnote
{As of the time of the meeting, June, 1999.  New discoveries since
then are discussed in the update section at the end of
this article.}.  I review 
current techniques for finding quasars at $z>5$
and the status of current optical surveys.  I compare
the spectra of known quasars with $z \approx 5$ with
the spectra of some recently discovered galaxies with $z>5$ to see
what we may expect in the future from surveys
for high redshift quasars and galaxies.  The prominent emission lines
of quasars should make them easier to detect and confirm 
spectroscopically than the $z>5$ galaxies discovered so far.


\end{abstract}

\section{Introduction and Background}

I am grateful to the Organizing Committee for the opportunity to speak
on searches for high-redshift quasars at this meeting.  Although I have not
worked directly with Hy, we have known each other for many years, and 
I have admired his research.  I think three qualities stand out: 1) his determination
and persistence at pushing telescopes and instruments to their limits, 2) his evident 
success at finding and studying successively more distant galaxies, and 3)
his cheerfulness and optimism, which I think inspired his students and
collaborators to be so successful over the many years of his career.  I 
have also
heard that Hy never liked quasars very much, so I feel honored that the topic
made it on the program.  Maybe the organizing committee decided it
was safe to do so, now that galaxies have overtaken quasars as the objects
of highest known redshift since 1997, when they displaced
quasars for the first time since 1965.

Turning to the subject of this paper, the goals of surveys for 
high-redshift quasars
include the observational determination of:

\begin{itemize}

\item The epoch of quasar formation

\item The role of dust obscuration

\item The relation of quasars to galaxy formation and evolution

\item The contribution of quasars to the ionization of the intergalactic medium
at high redshifts

\end{itemize}

According to the picture that quasars are powered by the accretion
of matter onto massive black holes in the nuclei of galaxies, the
formation and evolution of black holes, quasars, and galaxies are closely
related.  The evolution of the space density of luminous
quasars shows a strong
peak at lookback times of 0.8 to 0.85 (Fig. 1).  
The evolution of galaxies is very
much under debate, as we are hearing at this meeting, but it appears to
extend over a broader range of redshift.

\begin{figure}
\plotfiddle{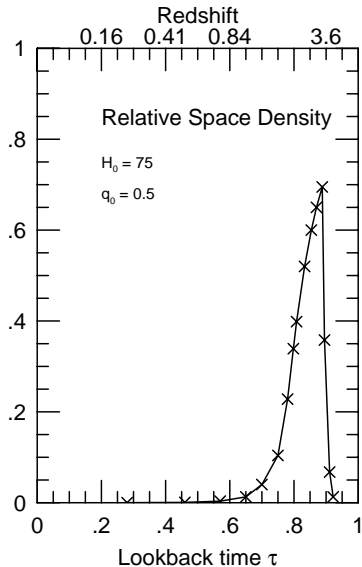}{3in}{0}{28}{28}{-100}{0}
\caption{A linear plot of space density vs. lookback time
for quasars with $M_C < -24.5$,
based on the results of Warren et al. (1994).  $M_C$ is the AB continuum
magnitude at the observed wavelength corresponding to Ly$\alpha$.}
\end{figure}

Among the questions we would like to address are: Which formed first, galaxies
(or at least parts of galaxies), or quasars? What roles do interactions and 
mergers play in the fueling of quasars? and How do we account for the 
chemical abundances in high-redshift quasars, whose emission-line spectra
look very similar to those at lower redshift?

We now have the observational capabilities to map the evolution of both
quasars and galaxies to redshifts beyond 6, when the universe was 5\% or
less of its present age.  Hubble Space Telescope and 8 - 10-m ground-based
telescopes have already enabled us to discover more objects at $z > 5$
than seemed possible just a few years ago.  Now the challenge is to carry
out sufficiently large, systematic, and quantifiable surveys to answer
the questions posed above.

In this paper I will discuss how to search for quasars at $z > 5$, 
concentrating on optical methods.  Next I will review the status
of current surveys and then compare the spectra of the quasars 
and galaxies with $z >\approx 5$ that have been found to date.
I will close by making some speculations based on the current
results and considering what we may expect in the future.

\section{How to Search for Quasars at $z> 5$?}

Since the discovery of the first quasar with $z > 4$ in 1987
(Warren et al. 1987),
more than 150 such objects have been found (G. Djorgovski
and R. McMahon, private communications), and
several search techniques are by now well developed.  They include:
\begin{itemize}
\item Multi-color imaging
\item Slitless Spectroscopy
\item Deep Radio Surveys
\item Deep X-Ray Surveys
\end{itemize}

The multicolor
technique pioneered by Warren et al. (1987) and now applied to 
several surveys (see, e.g., Warren et al. 1991, Irwin et al. 1991,
Kennefick et al. 1995, Djorgovski 1999) has contributed the
majority of the known $z > 4$ quasars.  It uses the large
$B - R$ color of quasars as a discriminant;  the 
presence of Ly$\alpha$ emission in the $R$ filter band combined
with the continuum depression caused by intervening Ly$\alpha$
absorption in the $B$ filter band makes quasars stand out
from cool stars and yields an effective search criterion.  The
addition of $I$ band photometry, which is an indicator of the continuum
level longward of Ly$\alpha$, aids in distinguishing 
high-redshift quasars from late-type stars.  The multicolor technique
was first used with photographic plates and Schmidt telescopes.
The subsequent development of
large format CCD cameras is now enabling wide-angle surveys
to significantly fainter magnitudes, as discussed in more detail
below.

At higher redshifts, e.g. $z > 5$, the same principle applies, with
Ly$\alpha$ shifting into the $I$ band.  However, as we shall see below,
it is critical to have observations in the $Z$ band, which is in the
continuum longward of Ly$\alpha$, to separate quasars from late-type
stars.

The slitless spectroscopy technique is an effective way
to discover high-redshift quasars through the direct detection
of Ly$\alpha$ emission in low-resolution slitless spectra
(Smith 1975, Osmer 1982) and has been used by Schmidt, Schneider,
and Gunn (SSG, 1995 and references therein) in their Palomar 
Grism Surveys.  The
selection effects and survey efficiencies are more straightforward
to model for slitless spectroscopy in many cases than 
they are for the multicolor technique.  The SSG survey is
one of the cornerstones of our knowledge of the evolution
of optically selected quasars at $z > 3$.

Deep radio surveys offer a different way to discover
high-redshift radio galaxies and quasars. A key property
of radio-selected objects is that they
are significantly less affected by surrounding
or intervening dust that blocks optical radiation.  Although
radio-loud quasars in general constitute about 10\% of
the total quasar population, the
availability of deep radio surveys that cover a significant
fraction of the sky offers the opportunity to find
significant numbers of radio-loud quasars and radio galaxies
at high redshift.

Current radio searches for high-redshift quasars make
use of deep surveys for flat-spectrum sources (e.g. Hook and
McMahon 1998) while searches for high-redshift radio galaxies
focus on steep-spectrum sources (e.g. van Breugel
et al. 1999, also this volume).  In both cases, subsequent
selection concentrates on optically faint objects to
help weed out objects of lower redshift. Hook and McMahon
(1998) select objects that are red in $B - R$ to enhance
the selection of high-redshift objects. van Breugel et al.
(1999) make use of the well-defined $K~-~z$ diagram for
radio galaxies and concentrate on objects with faint
$K$ magnitudes.  The discoveries of the radio-selected
quasar of highest known redshift, $z = 4.72$ by Hook and
McMahon (1998) and the radio galaxy of highest known redshift,
$z = 5.19$ by van Breugel et al. (1999) demonstrate the power of
their approaches.

It may be argued that X-ray emission is the key defining property
of the global quasar/AGN population and should be the primary
search criterion for discovering them.  However, until recently,
X-ray observatories have not had sufficient sensitivity to reach
the objects of highest redshift.  The ROSAT Deep Survey
(Hasinger et al. 1998) is an indicator of the advances
we may expect from CHANDRA (e.g., Mathur, this volume)
and XMM as they come into operation;
it has yielded the X-ray selected quasar of highest known
redshift, $z = 4.45$ (Schneider et al. 1998).  The requirements
are a limiting X-ray sensitivity of $\approx 10^{-15}$ erg
s$^{-1}$ cm$^{-2}$ (0.5 -- 2keV) and high accuracy, 
$\approx 1$ arcsec positions
(to avoid confusion with foreground objects).  Then follow-up
optical imaging and spectroscopy, presumably concentrating
on faint, red objects, should be an effective way of isolating
high-redshift objects.

To summarize, all approaches to find $z > 5$ quasars need to cover
a wide area on the sky to faint limiting sensitivities because the
objects are so rare.  Optical
surveys need to have wavelength coverage extending to 0.9$\mu$;
indeed all approaches benefit from such coverage.  Searches at
even higher redshifts will push the requirements into the infrared
($\lambda > 1\mu$).

\section{Status of Current Optical Surveys}

I would like to give a status report on two multicolor surveys my
collaborators and I are carrying out: the BTC40 and the BFQS.
I will also describe two other major optical
surveys, CADIS and SDSS, and note the important contributions they
are making to the search for quasars at $z > 5$.

The BTC40 is a large collaborative effort led by E. Falco and is aimed
at finding gravitational lenses and distant galaxy clusters as well
as high-redshift quasars.  It is an optical, multicolor survey that
makes use of the BTC camera and the CTIO 4-m telescope.  Julia
Kennefick is leading the quasar survey team, whose other members
include Alberto Conti, Richard Green, Pat Hall, Eric Monier, Malcolm Smith, and
myself.  BTC40 has imaging data at high galactic latitude for 40 deg$^2$
in the $B,V,I,Z$ bands and reaches to $\approx$ 25th magnitude at its
deepest limit.  All the imaging data are in hand and are being analyzed.
The goal is to select $z > 5$ quasar candidates to $I=22$.

The BFQS survey is being led by Pat Hall in collaboration with the other
team members mentioned above.  It also uses the BTC camera and reaches to $m_{lim} = 26.7$ over 7 deg$^2$.  It uses the $B,R,I$ bands and
is aimed at quasars with $3.3 < z < 5$ down to L* luminosities.
The BFQS imaging data are also in hand and are being analyzed.

Together, the BTC40 and BFQS surveys extend the multicolor technique
to fainter magnitudes and wider areas than have been covered before.

The CADIS (Calar Alto Deep Imaging Survey, Meisenheimer et al. 1998)
survey extends the multicolor concept to higher spectral
resolution and broader wavelength coverage by using 13 medium-band filters
over the 0.39 to 0.93$\mu$ range and 3 broad-band filters, 
including $K^\prime$.  The filters are designed to give both improved
sensitivity to the quasars being sought and improved rejection of
non-quasars, such as emission-line galaxies (the veto filter concept).
CADIS plans to cover 9 fields, each $10\arcmin \times 10\arcmin$ in
area.  To date the CADIS team (Wolf et al. 1999) have shown the
effectiveness of their approach by finding 6 quasars with 
$2.2 < z < 3.7$ to $R=22$ mag in one field, about 6 times the 
expected number.  The CADIS technique is more powerful than the
original multicolor approaches because of its increased spectral
resolution and greater number of filters.  On the other hand, it
requires more observing time per unit area on the sky to reach 
a given limit in sensitivity.  Thus, it is a complementary and valuable
addition to techniques for finding high-redshift quasars.  It will be
very important to see if the initial CADIS results on the surface density
apply to the additional fields in their survey.  Previous surveys have
been subject to considerable field-to-field fluctuations, and it will
be very interesting to see how the CADIS results come out.  Our knowledge
of the luminosity function of quasars to 22nd magnitude and fainter is
still rudimentary.

The SDSS (Sloan Digital Sky Survey, Gunn \& Weinberg 1995) 
will be the definitive survey
for high-redshift quasars (and also for quasar clustering) down to 20th
magnitude because of its large area coverage, 10,000 deg$^2$. SDSS
is expected to yield 10$^5$ quasars when completed.  Already the initial
results are very exciting. Fan et al. (1999) report the discovery
of 15 quasars with $z > 3.6$ in 140 deg$^2$, including quasars with
$z=4.9$ and $z=5.0$.  They extend the high-redshift record for quasars
for the first time since 1991\footnote{After the meeting, Fan et al.
(2000a) reported the discovery of a quasar at $z=5.03$.}.

\subsection{The Importance of the $Z$ band}

Initial spectroscopic results from the BTC40 survey confirm the importance
of using the $Z$ band for isolating $z > 5$ quasar candidates.  Unlike
the situation at $4<z<5$, where quasars can be selected with reasonable
efficiency solely on their large values of $B-R$, the analogous approach
using $V-I$ does not work for $z>5$ quasars.  Julia Kennefick
found that samples selected on the basis
of large $V-I$ are overwhelmed by late-type stars.  The $Z$ band is 
necessary because it is a measure of the continuum level at wavelengths
longward of the Ly$\alpha$ emission, and the behavior of late-type
stars and $z>5$ quasars is significantly different in, for example,
the $(I-Z)~vs.~(V-I)$ two-color diagram, as Kennefick's simulations
show (Fig. 2, left).  We are now in the process of selecting 
$z>5$ candidates for follow up with spectroscopic observations.
%Preliminary application of this approach to a 
%5 deg$^2$ area of the BTC40 sample produced 9 candidates for
%high-redshift quasars.  George Djorgovski (private communication)
%kindly obtained Keck spectra of 3 of them for us.  He indicates
%from a first look at the results suggests two objects are very
%late-type stars and the third is not clearly identified yet.

The importance of the $Z$ band is amply demonstrated 
in the SDSS results (Fig. 2, right),
where Fan et al. (1999) illustrate the stellar locus in the 
$(i^*-z^*)~vs.~(r^*-i^*)$ plane and the location of their $z>4.5$ quasars,
which are well separated from the stars.
Their results confirm that we now have well established techniques 
for finding the most distant quasars.

\begin{figure}
\plottwo{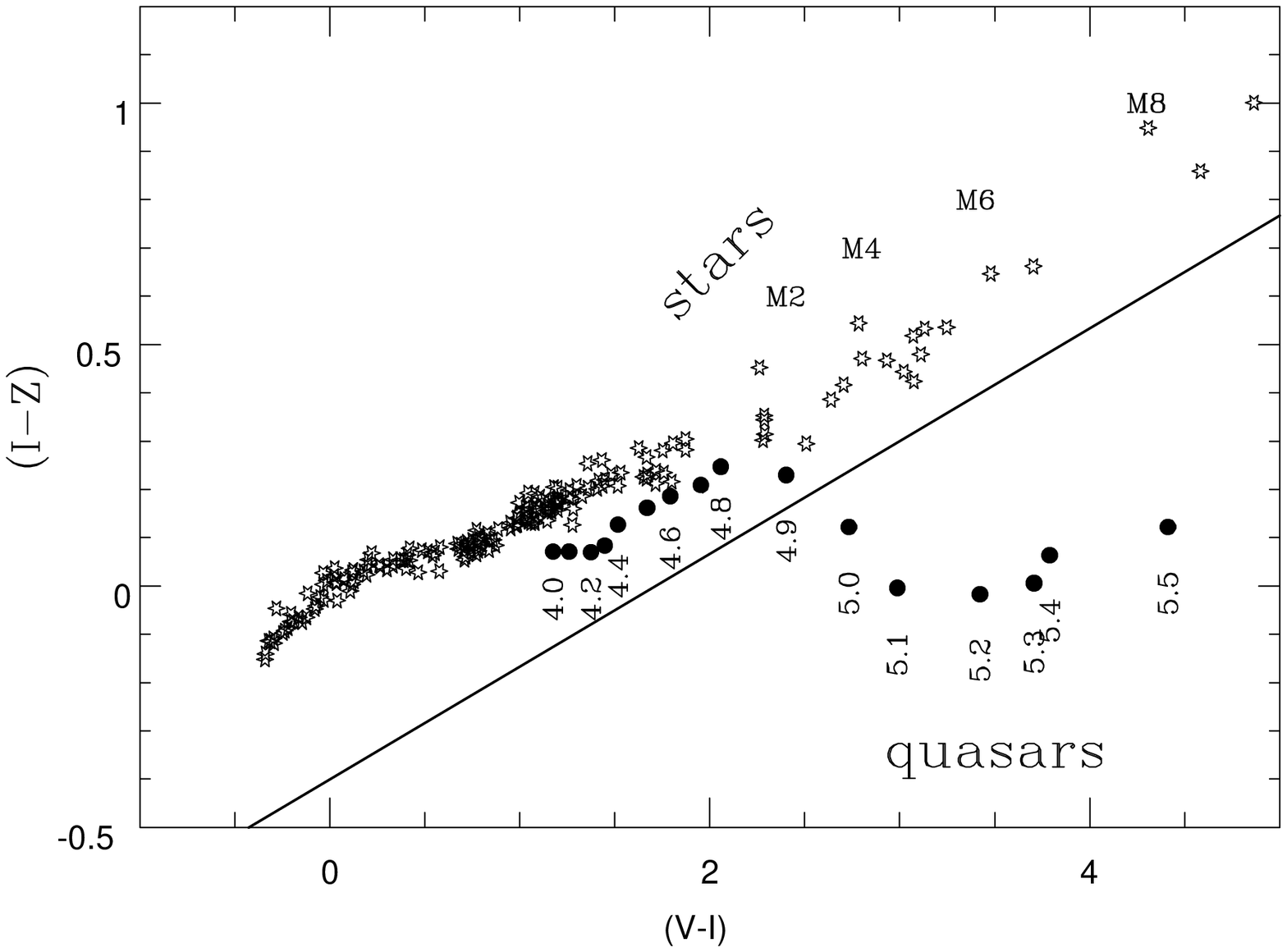}{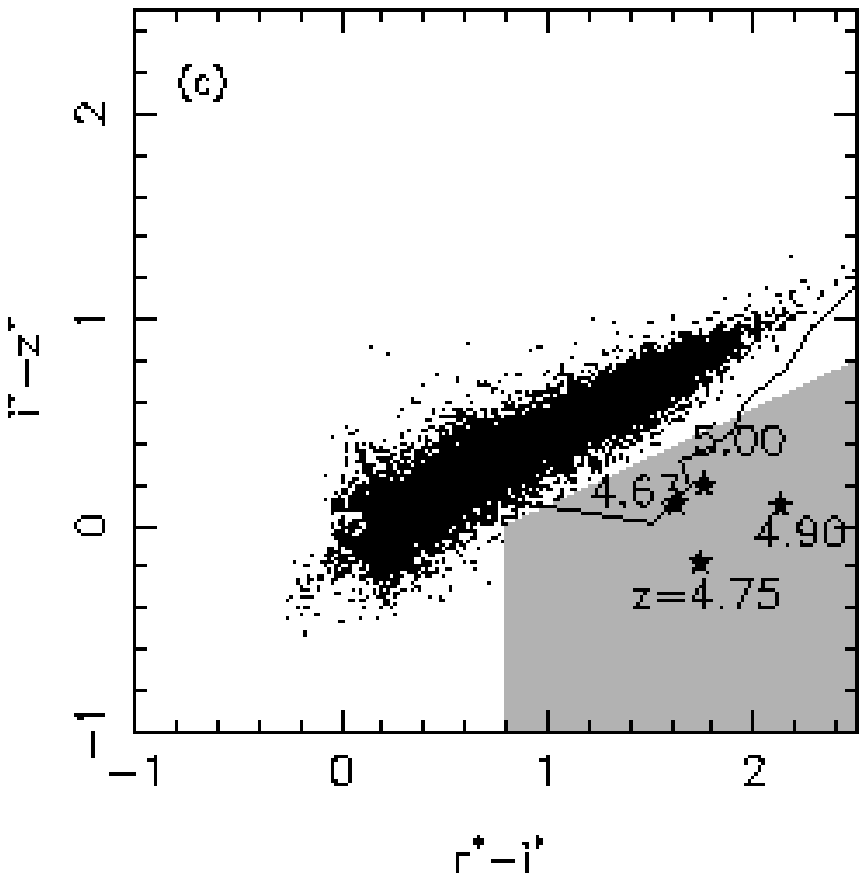}
\caption{(left) Simulation of stellar and quasar colors by
J. Kennefick. (right) The first SDSS results for stars and $z >4.5$
quasars (Fan et al. 1999)}
\end{figure}

\section{The Spectra of $z \approx 5$ Quasars Compared to 
$z > 5$ Galaxies}

It is of interest to note that the spectra of
the $z \approx 5$ quasars discovered to date (Fig. 3) show the 
characteristically strong emission features of
Ly$\alpha$, {\sc C IV, NV} and {\sc Si IV + O IV]} seen
in quasars at lower redshift and look remarkably similar
to them.  Furthermore, the spectra provide
information on the chemical abundances in the broad-line
region that have important implications for the evolutionary
history of the host galaxies of quasars (see review
by Hamann \& Ferland 1999).

\begin{figure}
\plottwo{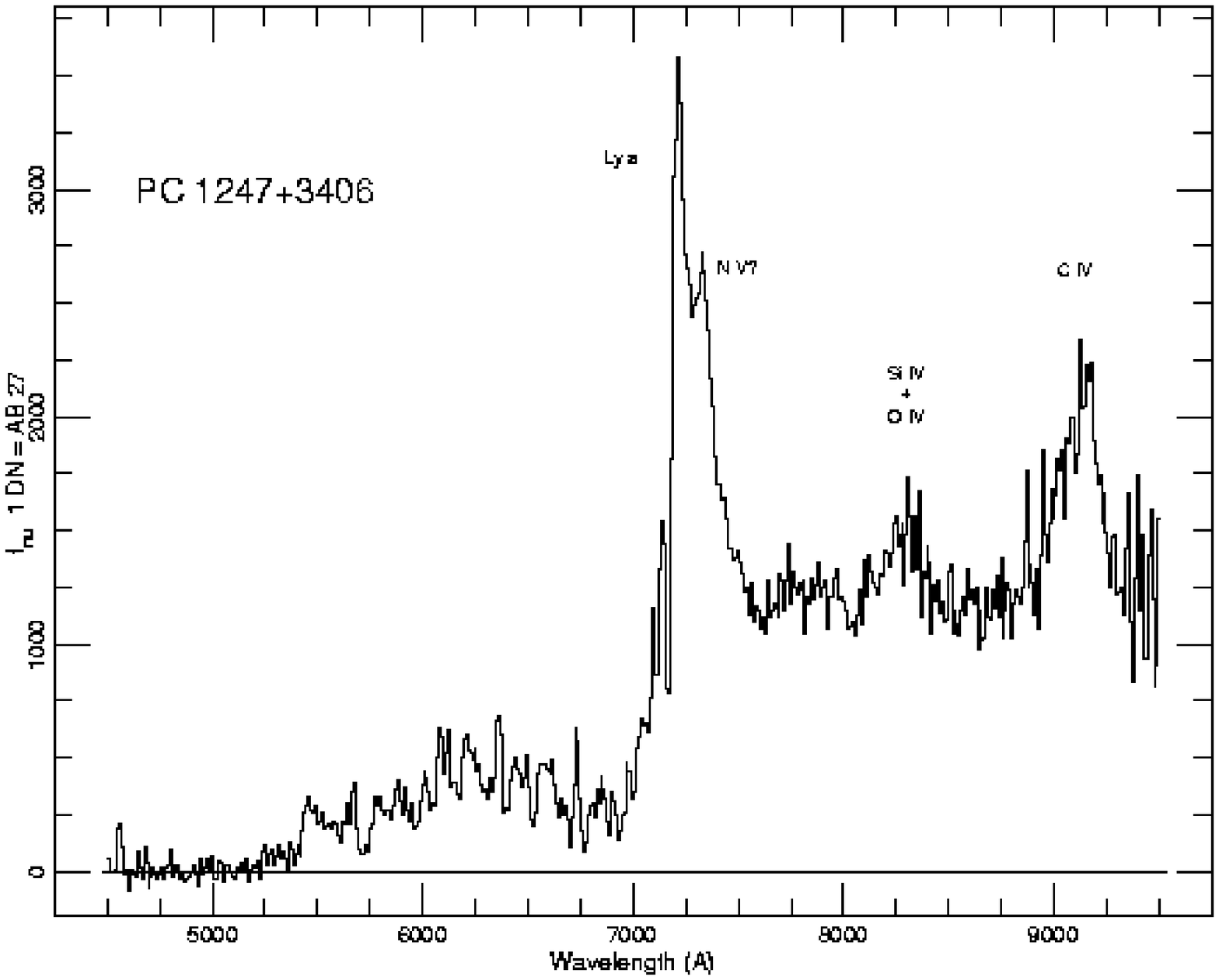}{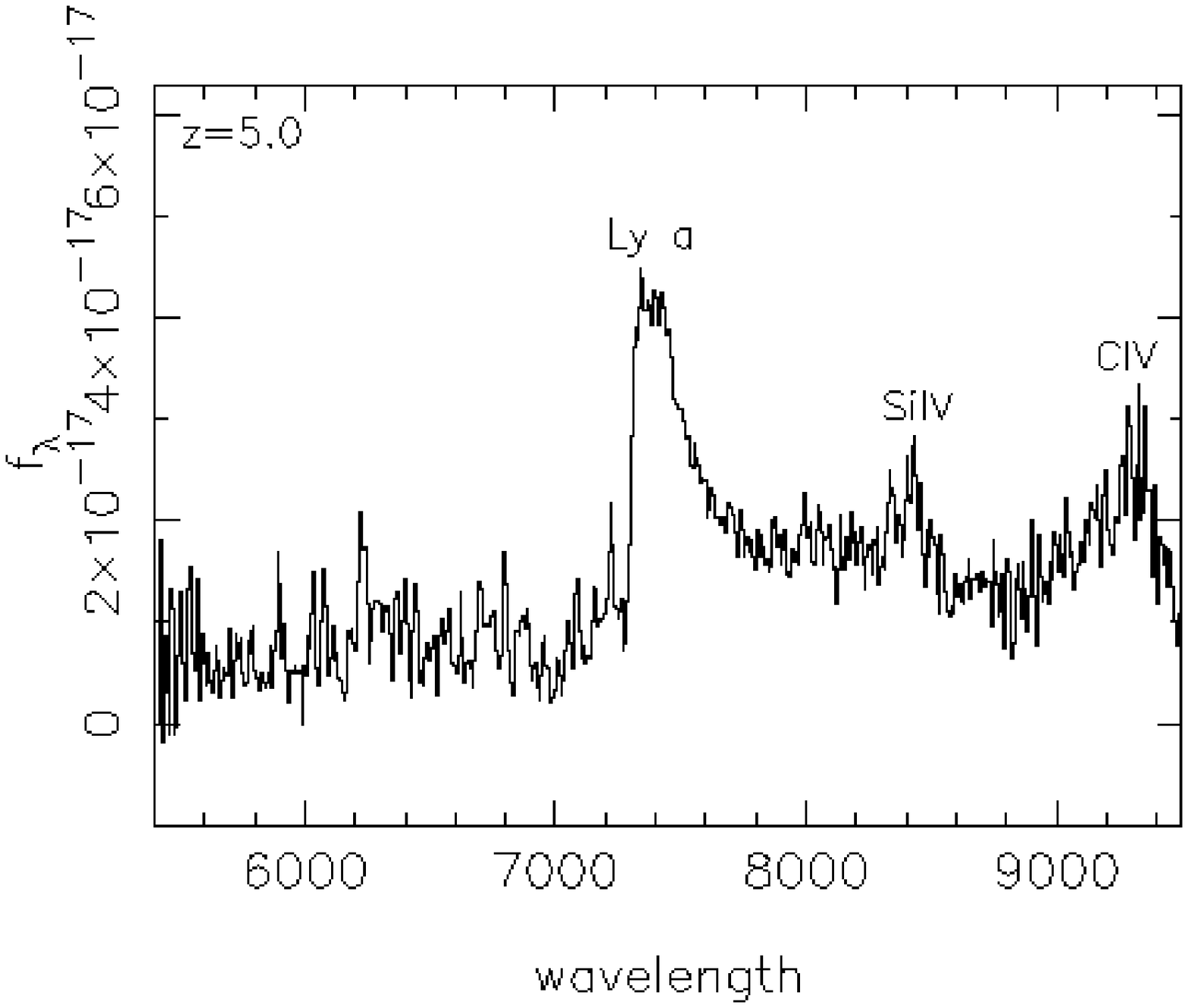}
\caption{Spectra of PC1247+3406, $z=4.9$ (Schneider et al. 1991)
and the SDSS $z=5$ quasar (Fan et al. 1999)}
\end{figure}

On the other hand, the spectra of the galaxies found so far
at $z > 5$ are very different from those of quasars (Figs.
4 and 5).  For example, the Dey et al. (1998) galaxy at 
$z=5.34$ has a very strong but narrow Ly$\alpha$ emission 
line on a weak continuum.  The Weymann et al. (1998) galaxy 
at $z=5.60$ and the van Breugel et al. (1999) radio galaxy show Ly$\alpha$
emission that is considerably weaker, while the Spinrad et al.
(1998) galaxy pair at $z=5.34$ has very weak, if any, emission and
is distinguished primarily by a break in the continuum.

\begin{figure}
\plottwo{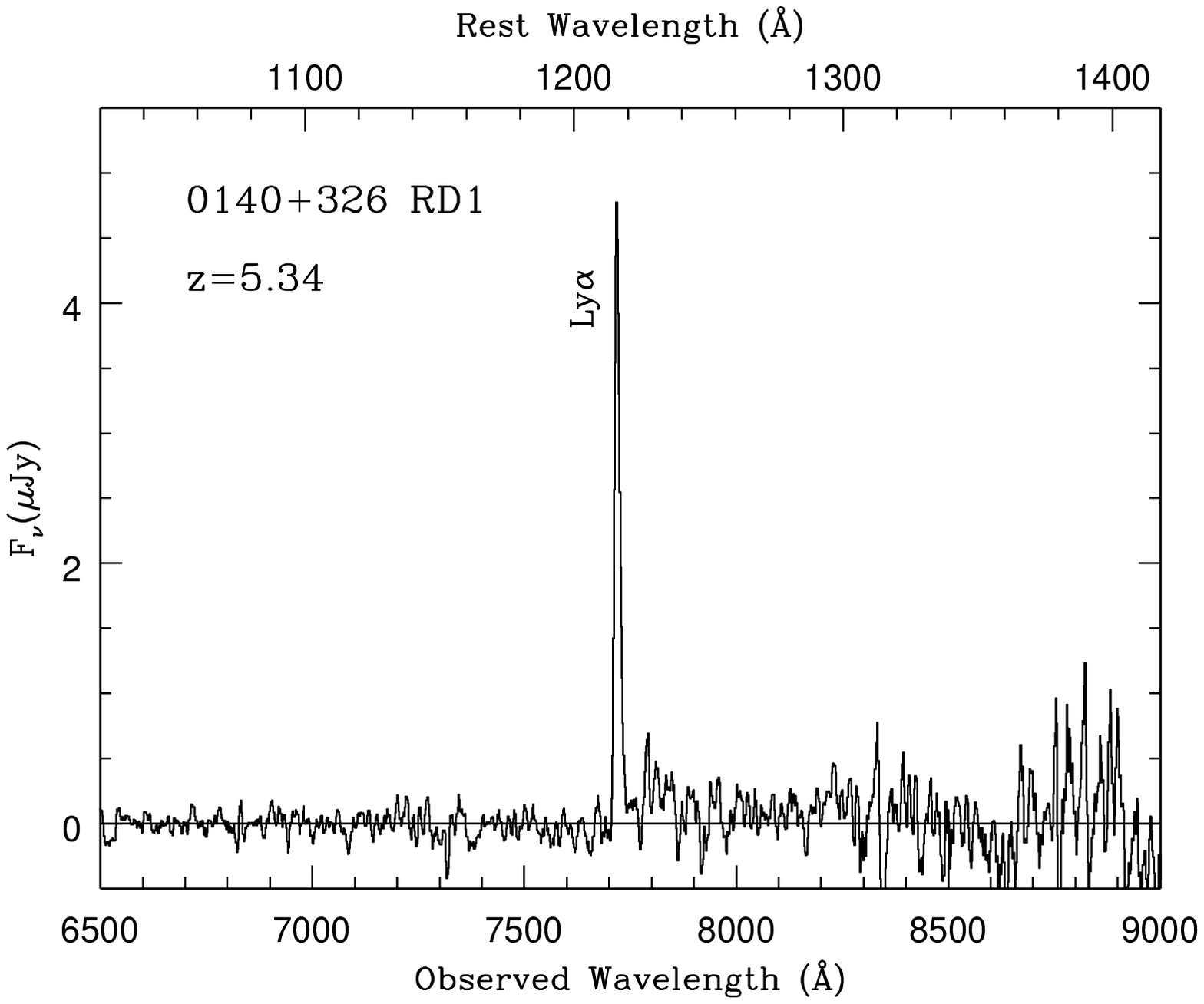}{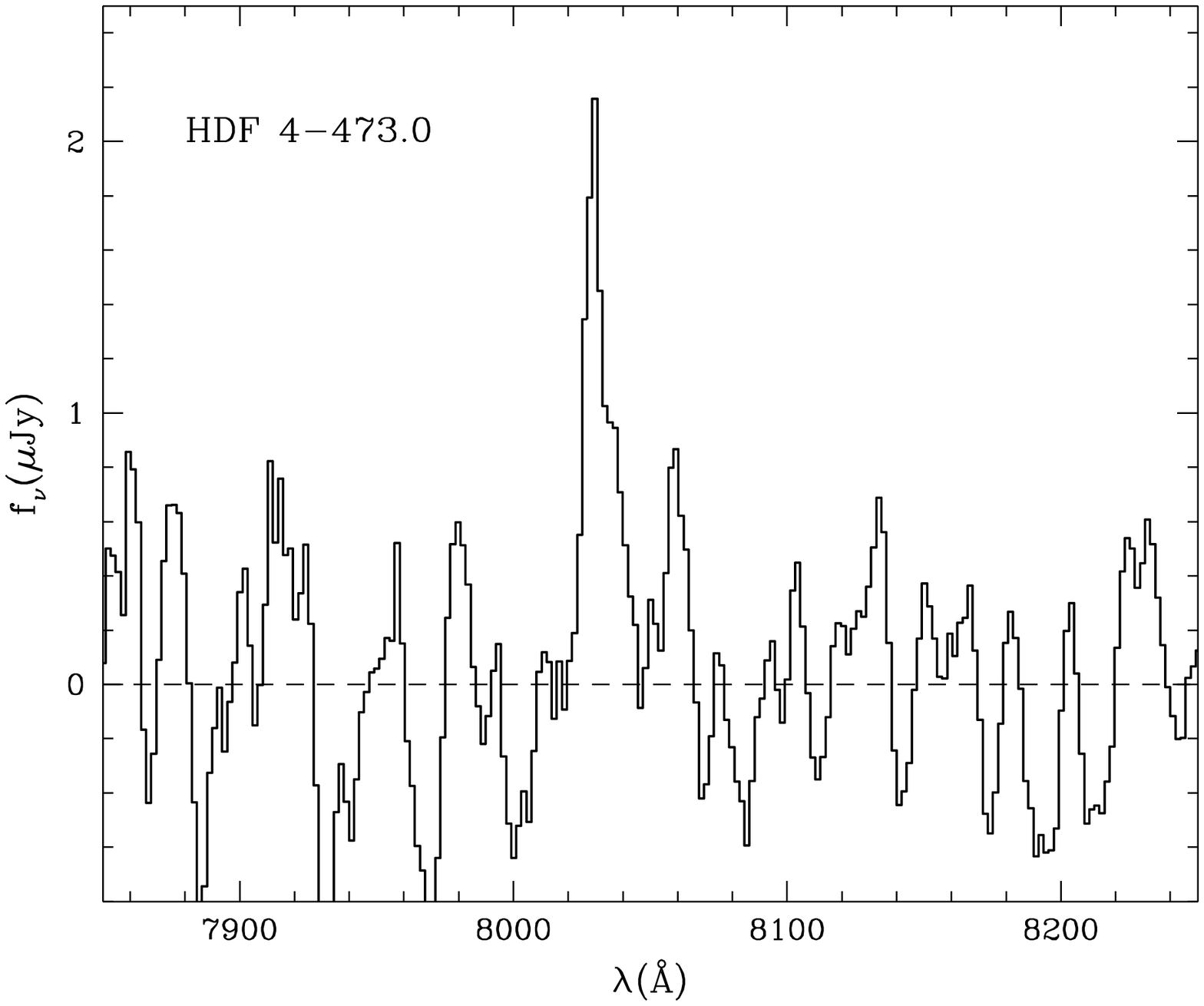}
\caption{Spectra of the Dey et al. (1998) galaxy at $z=5.34$ (left)
and the Weymann et al. (1998) galaxy at $z=5.60$ (right)}
\end{figure}

\begin{figure}
\plottwo{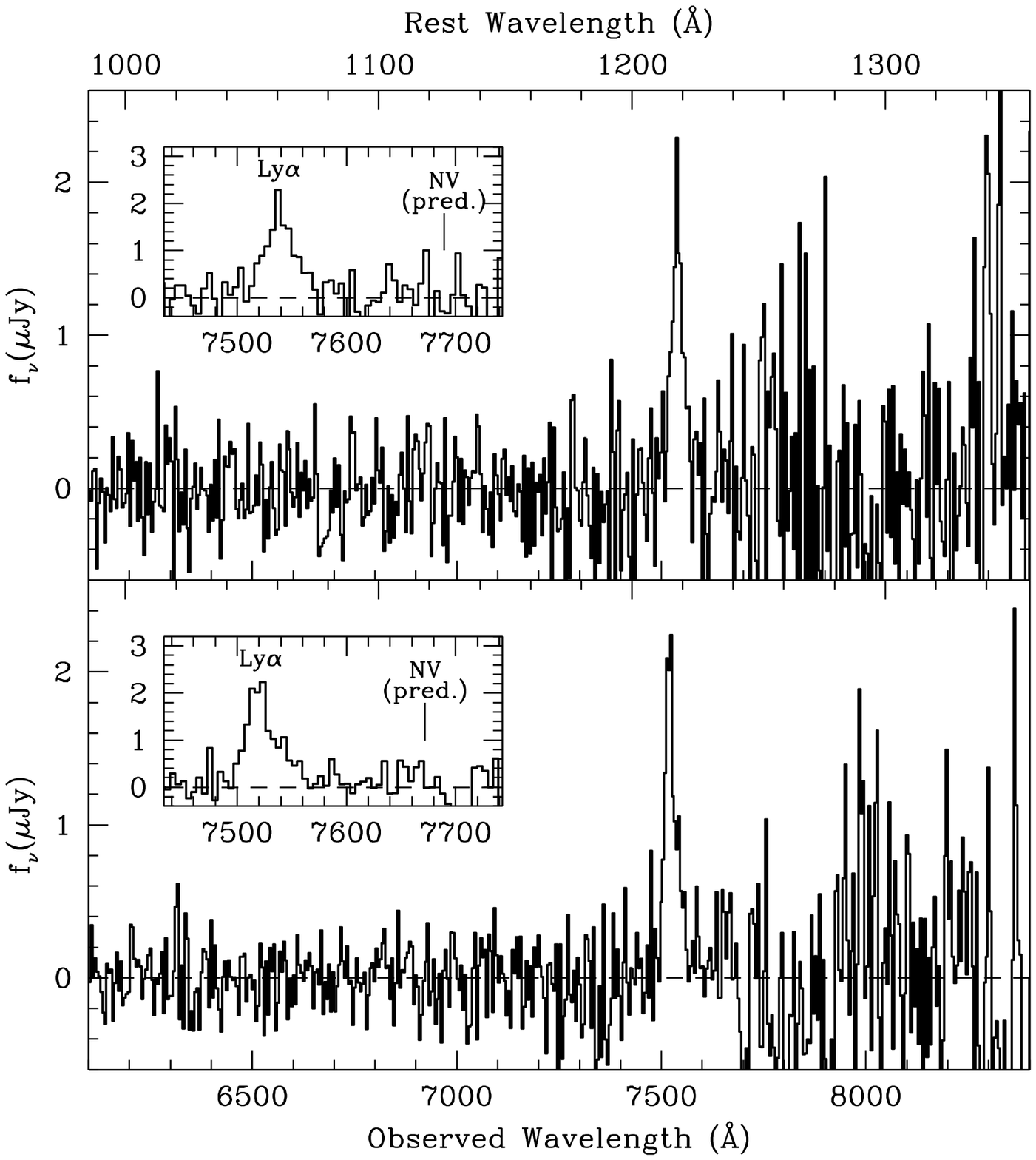}{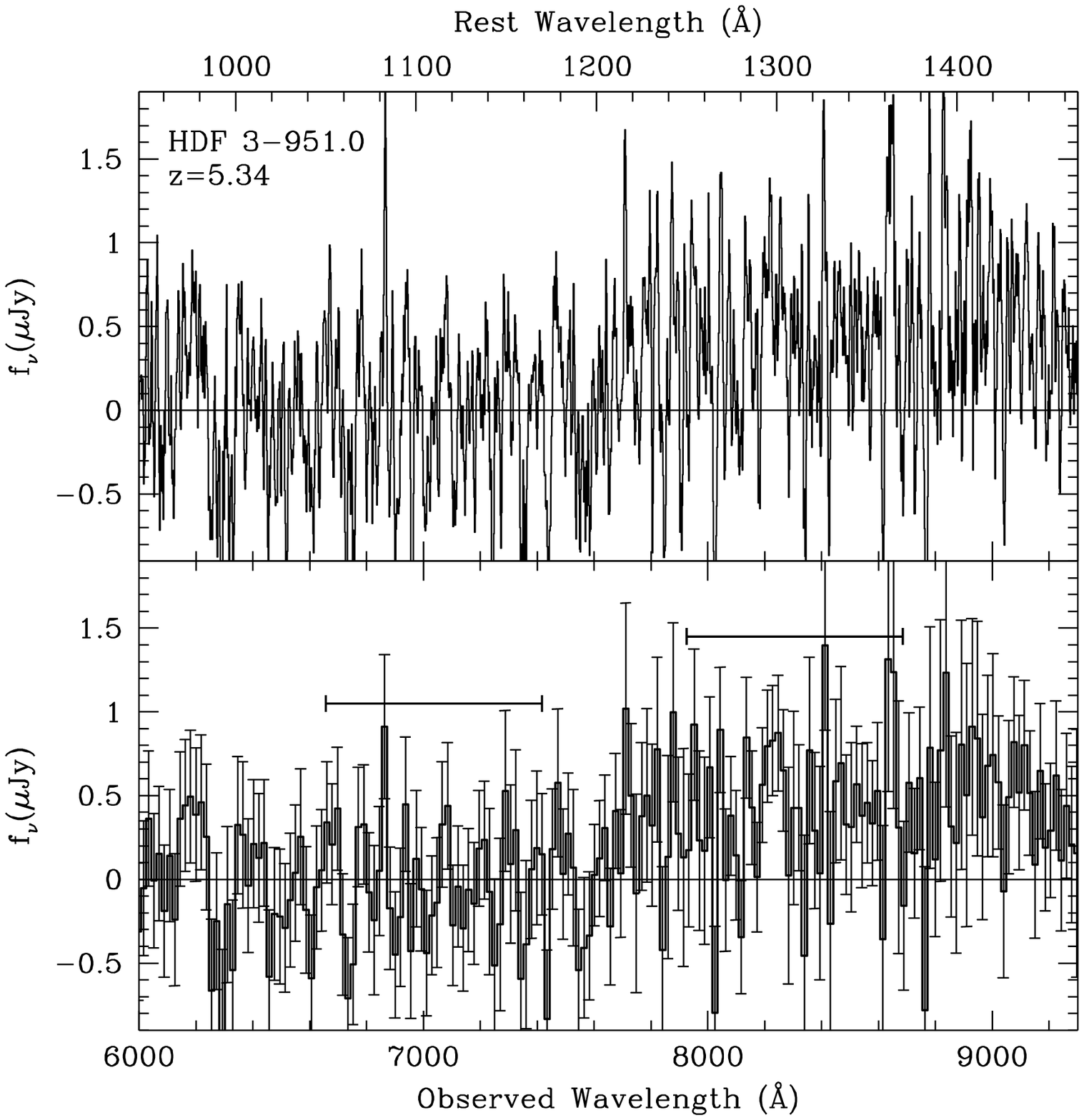}
\caption{Spectra of the van Breugel et al. (1999) radio galaxy at
$z=5.19$ (left) and the Spinrad et al. (1998) galaxy pair
at $z=5.34$ (right)}
\end{figure}

Let us consider for a moment the faintness of these galaxies and 
the large amount
of spectroscopic observing time that has been dedicated
by several groups of observers with the Keck telescopes to
searching for faint galaxies at high redshift.  It is surely
impressive that galaxies with $z>5$ are being found.  The numerous
discoveries exceed what I thought only a few years ago would occur.
But the weakness of the spectral features of the galaxies shown
in Figs. 4 and 5 compared to the strength of quasar emission lines
in Fig. 3 do make one wonder if quasars/AGNs with broad-line spectra
are truly very scarce at $z>5$.  Their spectra are much easier to
identify than those of the galaxies seen to date.  What is their
absence telling us?

Of course, it may just be too early to tell, and we must pursue
the spectroscopic follow up of the $z >5$ surveys described above.
In any case, the observational opportunities that
we now have to study the universe at $z >5$ are very exciting, and
we may expect continued important discoveries in the near future.
Let's meet again on the occasion of Hy's 70th birthday to see what
we have learned.

\section{Update, May, 2000}

Between June, 1999, when the meeting was held, and May, 2000,
more quasars with $z > 5$ have been discovered, including ones
at $z=5.27$ (Zheng et al. 2000),
at $z = 5.5$ (Stern et al. 2000) and at $z = 5.8$ (Fan et al. 2000b).
The $z=5.5$ object, which has $I_{AB} = 23.8$, is the faintest known 
quasar at
high redshift.  The one at $z=5.8$, with $z^* = 19.2$ on the AB system,
is very luminous.  All have the strong emission lines 
characteristic of quasars at lower redshifts, and all have
strong Ly$\alpha$ forest absorption at wavelengths shortward
of their Ly$\alpha$ emission lines. However, the universe is still
highly ionized at $z=5.8$. Fan et al. (2000b) report
that they have covered 600 deg$^2$ so far in their work on the
Sloan Digital Sky Survey and that the discovery of the 
$z = 5.8$ quasar is consistent with the expectations of
the Schmidt et al. (1995) parameterization of the decline
of the quasar luminosity function with increasing redshift.
They also note that for $z > 5.5$ it is difficult to distinguish
quasars from very cool stars and brown dwarfs on the basis of
the SDSS filters alone. Observations in the
near infrared, where the cool objects have redder energy
distributions than the quasars,
aid in the selection of the quasars.

\acknowledgements

I am grateful to the Organizing Committee for the invitation
to speak and for their financial support.  This work has
also been supported by the National Science Foundation
under Grant AST-9802658.  Figures 2b, 3
\& 5b are reproduced from the Astronomical Journal, and figures 4 \& 5a
are from the Astrophysical Journal (Letters), courtesy of the American
Astronomical Society and the authors. Arjun Dey, Xiaohui Fan, Donald
Schneider, Hyron Spinrad and Wil van Breugel are thanked for giving
permission to include their figures in this review article.

\end{document}